\documentclass[a4paper,11pt]{article}

\usepackage{amssymb}
\usepackage{makeidx}
\usepackage{moreverb}
\usepackage{algorithmic}
\usepackage{algorithm}
\usepackage{mathrsfs}
\usepackage{multicol}
\usepackage{amsthm}
\usepackage{amsfonts}
\usepackage{graphicx}
\usepackage{float}
\usepackage{amsmath,amssymb,latexsym,comment}
\usepackage{hyperref}
\usepackage[utf8]{inputenc}
\usepackage[small,bf,labelsep=period]{caption}

\newcommand{\triplesum}{\sum_{{\bf \kk} \in \KK }}
\newcommand{\kk}{\bf \gamma}
\newcommand{\KK}{\Gamma}

\newcommand{\A}{\mathbb{A}}
\newcommand{\B}{\mathbb{B}}
\newcommand{\R}{\mathbb{R}}

\newcommand{\N}{\mathbb{N}}
\newcommand{\Z}{\mathbb{Z}}
\newcommand{\M}{\mathbb{M}}
\newcommand{\C}{{\mathcal C}}
\newcommand{\Nn}{{\mathcal N}}
\newcommand{\SM}{{\mathcal S}}

\newcommand{\K}{{\mathscr K}}
\newcommand{\Per}{{\mathscr P}}
\newcommand{\Ker}{\mathbb{K}}
\newcommand{\fracd}[2]{\displaystyle{\frac{{\displaystyle{#1}}}{{\displaystyle{#2}}}}}
\newcommand{\fracsum}[2]{#1 \; / \; #2 }

\begin{document}



\title{Hiding data inside images using orthogonal moments}
\author{Anier Soria-Lorente$^{1}$, Stefan Berres$^{2}$ and Ernesto Avila-Domenech$^{1}$\\[.2cm]
	$^1$ Tecnology Department\unskip, 
	University of Granma\unskip, Bayamo\unskip, 85100\unskip, Granma\unskip, Cuba\\
	$^2$ Departamento de Ciencias Matem\'{a}ticas y F\'{\i}sicas\\
	Facultad de Ingenier\'{i}a\unskip,  Universidad Cat\'olica de Temuco\unskip, Temuco, Chile\\
	asorial@udg.co.cu, eadomenech@gmail.com, sberres@uct.cl}
\date{\emph{(\today)}}
\maketitle

\begin{abstract}
In this contribution we propose a novel steganographic method based on several orthogonal polynomials and their combinations. The steganographic algorithm embeds a secrete message at the first eight coefficients of high frequency image. Moreover, this embedding method uses the Beta chaotic map to determine the order of the blocks where the secret bits will be inserted. In addition, from a 128-bit private key and the steps of a cryptography algorithm according to the Advanced Encryption Standard (AES) to generate the key expansion, the proposed method generates a key expansion of 2560 bits, with the purpose to permute the first eight coefficients of high frequency before the insertion. The insertion takes eventually place at the first eight high frequency coefficients in the transformed orthogonal moments domain. Before the insertion of the message the image undergoes a series of transformations. After the insertion the inverse transformations are applied to the original transformations in reverse order. The experimental work on the validation of the algorithm consists of the calculation of the Peak Signal-to-Noise Ratio (PSNR), the Universal Image Quality Index (UIQI), the Image Fidelity (IF), and the Relative Entropy (RE), comparing the same characteristics for the cover and stego image. The proposed algorithm improves the level of imperceptibility and security analyzed through the PSNR and RE values, respectively.

\medskip


\medskip


\medskip

\end{abstract}

\section{Introduction}

Nowadays, in the era of the so-called {\em Internet of Things} and ubiquitous computing,
all kind of communications is becoming to be strongly connected to the Internet, which has definitively become a backbone of the infrastructure of the modern world. This fact made it possible that the information transits by means of dissimilar communication channels, being used in considerable applications in science, in engineering, and in the industry~\cite{SRA}. Being a means of efficient communication, the Internet becomes a vulnerable tool for the information it carries, which can be acceded in many instances by non-authorized people, as well as consulted, modified and even destroyed. However, the great quantity of transmitted, potentially sensitive information requires sophisticated techniques of protection~\cite{ASoriaRP}. 

Most frequently secure communication is achieved by the method of encryption~\cite{SubhedaraSH}. Nevertheless, all electronic communications are being continuously and automatically monitored by both private and state-owned intelligent systems that have an enormous computer power. In particular, every transmission of cipher-text calls the attention of any of these systems and certainly is chosen to be analyzed, among others, by competitors and any sort of opposing forces. The use of electronic transmission media requires a method that calls less attention of the supervisory automatic systems. Modern Steganography offers a level of service that includes privacy, authenticity, integrity, and confidentiality of the transmitted data~\cite{SoriaSCN}. 

Steganography is the science or art of secret communication between two sides that attempts to conceal the existence of the message, such that the secrete data remains imperceptible to any adversary~\cite{AAwad}. Steganography methods embed the secret data in an appropriate and innocent multimedia carrier. The media with or without hidden information are called stego-media and cover-media, respectively. Often used media are text~\cite{gupta2018novel}, image~\cite{liao2018medical}, video~\cite{balu2018secure} and audio files~\cite{xin2018adaptive}. In
steganography, there are two common methods of embedding data, which can be classified into two categories, namely spatial domain and frequency-domain methods, see \cite{SoriaSCN} for more details. In the spatial domain method, the secret message is directly inserted into the least significant bit (LSB) of image pixels~\cite{hussain2018image}. In the frequency domain, however, the cover image is first transformed from the spatial to a frequency domain using some methods such as orthogonal moments, discrete wavelet transform or discrete cosine transform (DCT), then the secrete message is embedded in the transformed coefficients, and finally the data are transformed back from the frequency-domain to the spatial domain~\cite{SoriaSCN}. 

In spite of their scare use in steganography, orthogonal moments count with a large trajectory in image processing. The pioneering work on moment invariants was started by Hu (1962), who introduced this concept for pattern recognition~\cite{Hu}. Then, around two decades later, Teague~\cite{Teague} proposed continuous orthogonal polynomials as basis functions to calculate continuous moments such as Legendre or Zernike moments, thus demonstrating that images can be reconstructed with these orthogonal moments. However, these moments usually involve two kinds of inherent errors in digital images with a high computational cost. These error are (a) the discrete approximation of the continuous integrals, and (b) the transformation of the image coordinate system into the domain of the polynomials, see~\cite{Bayra}.

In order to solve this problem, discrete orthogonal moments such as Krawtchouk, Tchebichef, Hahn, Charlier and Meixner moments have been successfully introduced within the field of image analysis~\cite{jahid2018image,li2018tchebichef,sayyouri2016image,zhang2018medical}. The implementation of these moments does not require any numerical approximation since the basis set is orthogonal in the discrete domain of the image coordinate space~\cite{Bayra,Rahma}.

In steganography field, there are a couple of contributions that use orthogonal moments. Elshoura and Megherbi~\cite{Elshoura1} proposed two high capacity information hiding schemes in the frequency domain, in which large amounts of information can be hidden in gray level images with high transparency using Tchebichef moments. Whereas in~\cite{Elshoura2}, the authors presented a secure high capacity image information hiding scheme where two completely separate arbitrary full-scale gray level images, one hidden information image and one authentication watermark image are embedded hidden in the Tchebichef moments of a carrier image with very high imperceptibility.

Our main goal is to obtain a steganography system with a higher level of imperceptibility keeping up to an acceptable degree of security at the same time. Therefore, we introduce the idea of the use of some discrete orthogonal moments, which are powerful tools for processing images in the area of image analysis, but we now propose to apply this tool systematically for the purpose of stegonagraphy. Regarding the great variety of orthogonal moments, we focus principally on the discrete Krawtchouk, Tchebichef, Hahn, Charlier and Meixner orthogonal polynomials, as well as the $q$-Krawtchouk, $q$-Hahn, $q$-Charlier and $q$-Meixner orthogonal polynomials.

In this contribution, we describe a steganographic algorithm that embeds a secret message in the first eight high frequency coefficients of a given image. These coefficients are calculated by the orthogonal moments mentioned above and their combinations. In addition,  a 128-bit private key is used, which generates a binary sequence to permute the first eight high-frequency coefficients before insertion.
This strategy takes into account that modern steganographic techniques follow the Kerckhoffs's principle, according to which the opponent knows the technique used to hide the embedded message, and the security of the stego system is based only on the choice of hidden information shared between the sender and the receiver, called stegokey~\cite{RossiLGC}. Furthermore, this embedding method uses the chaotic map Beta \cite{zahmoul2017image} to determine the order of the blocks where the secret bits will be inserted, since this system is characterized by a pseudo-random behavior and a high sensitivity to initial conditions.

The structure of this paper is the following: In Section 2 we presented the literature survey. In Section 3, we introduce some preliminary results which will be very useful in the research presented. In Section 4, we describe the proposed embedding and extraction algorithms. Finally, in Section 5, we show the experimental results.

\section{A review of related works}
Several state-of-the-art steganographic research on hiding images in a DCT domain were reported in literature at recent years. In \cite{singh2012security}, the authors proposed a new robust steganography algorithm based on discrete cosine transform, the Arnold transform and chaotic systems. Thereby, the chaotic system is used to generate a random sequence to be used for spreading data in the middle frequency band DCT coefficient of the cover image. Moreover, the security is increased by scrambling the secret data using an Arnold Cat map before embedding. Mali et al. in \cite{mali2012robust} presented a robust and secured method of embedding a high volume of text information in digital cover images without incurring any perceptual distortion. Moreover, this method is robust against intentional or unintentional attacks such as image compression, tampering, resizing, filtering and Additive White Gaussian Noise. For the selection of the embedding locations in the frequency domain the Image Adaptive Energy Thresholding is used.
Then, in \cite{abdul2013secured} the authors present an elegant steganographic method at enhancing the reliability of the Mali et al.'s algorithm by overcoming the problem of misidentified blocks. To do so, an embedding map is adopted to indicate the location of the blocks. This means that some regions of the image are exploited for hiding data while some others are used for hiding the embedding map. In addition, the blocks, in which the data is concealed, are determined according to Mali et al.'s algorithm. In \cite{lin2014data}, a steganographic scheme based on the varieties of coefficients of the discrete cosine transformation of an image was proposed. Here, the authors use a integer mapping to implement the DCT, whereas that in \cite{karri2015steganographic}, a novel domain separation technique is proposed which is based on randomization of DCT kernel matrix. Habib et al. in \cite{habib2015enhancement} presented an interesting DCT steganographic method that spreads out randomly the secret bits within the cover image using chaos. Here, a digital chaotic generator based on two perturbed PWLCM is used to generate a binary stream, which is used to determine the positions of DCT coefficients in which the secret data is embedded. In \cite{el_rahman2016comparative}, the authors proposed a steganographic tool based on DCT, which is implemented to hide confidential information about a nuclear reactor, using the sequential embedding method in the middle frequency. Saidi et al. in \cite{saidi2017new} proposed a novel steganographic scheme based on chaotic map in the DCT domain, which applies the DCT on the cover image and scans the AC coefficients in a zigzag form the least significant to the most significant (inverse zigzag scanning). Here, the embedding$/$extracting process depends on a piecewise 
linear chaotic map, where its initial condition$/$control parameters are adopted as secret keys of the designed scheme. The authors of \cite{rachmawanto2017secure} proposed a combination of DCT steganography and cryptography using the one-time pad or Vernam cipher implemented on a digital image. In \cite{rabie2017high} a new discrete cosine transform approach for color image steganography is proposed. It implements a global adaptive-region embedding scheme that allows for extremely high embedding capacities while maintaining enhanced perceptibility.

In \cite{attaby2017data} the authors proposed a novel steganography technique of transform domain JPEG that provides high embedding performance while introducing minimal changes in the cover carrier image, thus maintaining minimum detectability against blind steganalysis schemes. 
This algorithm, named DCT-M3, uses the modulo  of the difference between two DCT coefficients to embed two bits of the compressed form of the secret message. In addition, Rabie et al. 
\cite{rabie2017toward} proposed a novel approach for color image steganography in the discrete cosine transform domain, that promotes an optimal embedding capacity while improving the stego image quality. This proposed approach is based on the observation that the space reserved for embedding the secret data varies with the statistical characteristics of the cover image and exploits a quadtree adaptive-region embedding scheme to individuate good cover image segments, in relation to the correlation of pixels, for embedding the secret information. 
Recently, in \cite{SoriaSCN}, Soria and Berres proposed a novel steganographic method based on the compression standard according to the Joint Photographic Expert Group and an Entropy Tresholding technique. This scheme uses one public key and one private key to generate a binary sequence of pseudo random numbers that indicate where the elements of the binary sequence of a secret bits will be inserted. Moreover, this algorithm improves the level of imperceptibility.
Then, in \cite{chowdhuri2018secured} Chowdhuri et al. presented a novel steganographic scheme, in which a weighted matrix is designed for highly compressed color images through a discrete cosine transform in order to maintain a good balance between payload and imperceptibility. After the color cover image is partitioned into three color channels (YCbCr), then then DCT coefficient matrix is obtained from each $(8\times 8)$ image blocks of from each YCbCr channel separately. Using a pre-determined quantization table, a quantized DCT coefficient is obtained from each block. Next, the AC coefficients, except 0, are collected from quantized DCT coefficients. The collection of AC components is controlled by a 128 bits shared secret key, which is used to generate a 512 bits binary stream by using SHA-512. Then, a series of $(3\times 3)$ original matrices are formed to hide secret data. Finally, a predetermined weighted matrix is employed to select the embedding position within a $(3\times 3)$ coefficient matrix of a cover image through the sum of the entry-wise multiplication operation.

\section{Preliminaries results}

In this section, a systematic representation of orthogonal polynomials is given that are later used for the domain mappings. 

\subsection{Discrete orthogonal polynomials}
The $n$th-order discrete orthogonal polynomials are those polynomials 
that satisfy the orthogonality condition~\cite{Nikiforov,Nikiforov1}
\begin{equation}
\sum_{x=0}^{\Nn}P_{m}(x)P_{n}(x)w^{\lambda}(x)=d^2_{n}(\lambda)\delta_{m,n},\quad \lambda=1,\ldots,5,
\label{defort}
\end{equation}
for the weight function $w^{\lambda}(\cdot)$ and the squared norm $d^2_{n}(\lambda)$, with the Kronecker delta function $\delta_{m,n}$, $\Nn\in\Z^{+}$ and $0\leq m,n\leq \Nn$.

Specific values for the parameters used in \eqref{defort}, namely for $\Nn$, $w^{\lambda}(\cdot)$, $d^2_{n}(\lambda)$, $\lambda=1,\ldots,5$, corresponding to the Krawtchouk $K_{n}^{p,N}(x)$ with $0<p<1$, Tchebichef $t_{n}^{N}(x)$, Hahn $H_{n}^{\alpha,\beta,N}(x)$ with $\alpha,\beta\geq -1$, Charlier $C_{n}^{\alpha}(x)$ with $\alpha > 0$ and Meixner $M_{n}^{\beta,\gamma}(x)$ with $\beta>0$, $0<\gamma< 1$, polynomials, respectively, are given in the Tables \ref{tableDOP}--\ref{tableCM}.


\begin{table}[ht!] 
	\caption{Characterization of the Krawtchouk, Tchebichef and Hahn polynomials.}
	\centering
	\renewcommand{\arraystretch}{2.0}
	\begin{tabular}{cccc} 
		\hline
		$P_n(x)$ & $K_{n}^{p,N}(x)$\: $(\lambda=1)$
		&$t_{n}^{N}(x)$\: $(\lambda=2)$
		& $H_{n}^{\alpha,\beta,N}(x)$\: $(\lambda=3)$ \\ \hline 
		$\Nn$ & $N$ & $N-1$ & $N-1$\\ 
		$w^{\lambda}(x)$ & $\binom{N}{x}p^x(1-p)^{N-x}$ & 1 & $\frac{(\alpha+1)_x(\beta+1)_{N-x}}{(N-x)!x!}$\\ 
		$d^2_{n}(\lambda)$ & $(-1)^n\frac{n!}{(-N)_n}\left(\frac{1-p}{p}\right)^n$ & $(2n)!\binom{N+n}{2n+1}$ & $\frac{(-1)^nn!(\beta+1)_n(\alpha+\beta+n+1)_{N+1}}{(-N)_n(2n+\alpha+\beta+1)N!(\alpha+1)_n}$\\ 
		$\alpha_n^N$ & $(Np-2np+n-x)\sqrt{\frac{(1-p)(n+1)}{p(N-n)}}$ & $1$ & $1$\\ 
		$\beta_n^N$ & $p(n-N)$ & $\frac{2x-N+1}{n}\sqrt{\frac{4n^2-1}{N^2-n^2}}$ & $A_n(x)\sqrt{\frac{d^2_{n-1}(3)}{d^2_{n}(3)}}$\\ 
		$\gamma_n^N$ & $\frac{n(1-p)^2}{p}\sqrt{\frac{(n+1)n}{(N-n)_2}}$ & $\frac{n-1}{n}\sqrt{\frac{2n+1}{2n-3}}\sqrt{\frac{N^2-(n-1)^2}{N^2-n^2}}$ & $-B_n(x)\sqrt{\frac{d^2_{n-2}(3)}{d^2_{n}(3)}}$\\ 
		$\K_{n}^{\lambda,N}(x)$ & $K_{n}^{p,N-1}(x)\sqrt{\frac{w^{1}(x)}{d^2_{n}(1)}}$ & $\frac{t_{n}^{N}(x)}{\sqrt{d^2_{n}(2)}}$ & $H_{n}^{\alpha,\beta,N}(x)\sqrt{\frac{w^{3}(x)}{d^2_{n}(3)}}$
		\\ \hline
	\end{tabular}
	\label{tableDOP}
\end{table}

\begin{table}[ht!] 
	\caption{Characterization of the Charlier and Meixner polynomials.}
	\centering
	\renewcommand{\arraystretch}{2.0}
	\begin{tabular}{cccc} 
		\hline
		$P_n(x)$ & $C_{n}^{\alpha}(x)$\: $(\lambda=4)$
		&$M_{n}^{\beta,\gamma}(x)$\: $(\lambda=5)$\\ \hline 
		$\Nn$ & $N$ & $N$\\ 
		$w^{\lambda}(x)$ & $\frac{e^{-\alpha}\alpha^x}{x!}$ & $\frac{\gamma^x\Gamma(\beta + x)}{x!\Gamma(\beta)}$\\ 
		$d^2_{n}(\lambda)$ & $\frac{n!}{\alpha^n}$ & $\frac{n!(\beta)_n}{\gamma^n(1-\gamma)^{\beta}}$\\ 
		$\alpha_n^N$ & $1$ & $1$\\ 
		$\beta_n^N$ & $\frac{\alpha -x+n-1}{\alpha}\sqrt{\frac{\alpha}{n}}$ & $-\frac{x-x\gamma-n+1+\gamma\beta+\gamma-\gamma n}{\gamma}\sqrt{\frac{\gamma}{n(\beta+n-1)}}$\\ 
		$\gamma_n^N$ & $-\sqrt{\frac{n-1}{n}}$ & $-\frac{(n-1)(n-2+\beta)}{n(n-1+\beta)}$\\ 
		$\K_{n}^{\lambda,N}(x)$ & $C_{n}^{\alpha}(x)\sqrt{\frac{w^{4}(x)}{d^2_{n}(4)}}$ & $M_{n}^{\beta,\gamma}(x)\sqrt{\frac{w^{5}(x)}{d^2_{n}(5)}}$\\ \hline
	\end{tabular}
	\label{tableCM}
\end{table}

The computation of the discrete orthogonal moments by using $K_{n}^{p,N}(x)$, $t_{n}^{N}(x)$, $H_{n}^{\alpha,\beta,N}(x)$, $C_{n}^{\alpha}(x)$ and $M_{n}^{\beta,\gamma}(x)$ presents numerical fluctuations~\cite{BarmakJ,MukundanOng,WuYan,Yap}.
Therefore a more stable version of them should be used. A normalized and weighted version of the discrete polynomials can be defined by $\K_{n}^{\lambda,N}(x)$, see Tables \ref{tableDOP}--\ref{tableCM}.

Indeed, $\K_{n}^{\lambda,N}(x)$, with $\lambda=1,\ldots,5$, satisfy the following recurrence relation~\cite{HLiao} for $\lambda=1,\ldots,5$,
\begin{equation*}
\alpha_n^N\K_n^{\lambda,N}(x) = \beta_n^N\K_{n+2\delta_{\lambda,1}-1}^{\lambda,N}(x)+\gamma_n^N\K_{n+\delta_{\lambda,1}-2}^{\lambda,N}(x),\quad n \geq 2-\delta_{\lambda,1},
\end{equation*}
where the coefficients $\alpha_n^N$, $\beta_n^N$ and $\gamma_n^N$ are given in Table \ref{tableDOP}. In addition, $A_n(x)$ and $B_n(x)$ 
(compare fourth column of Table \ref{tableDOP}) are given by
\begin{equation*}
A_n(x) = 1 + B_n(x) 
-x\fracd{(2n+\alpha+\beta+1)(2n+\alpha+\beta+2)}{(n+\alpha+\beta+1)(n+\alpha+1)(N-n)},
\end{equation*}
and
\begin{equation*}
B_n(x) =\fracd{n(n+\beta)(N+n+\alpha+\beta+1)}{(2n+\alpha+\beta)(n+\alpha+\beta+1)}\fracd{2n+\alpha+\beta+2}{(n+\alpha+1)(N-n)}.
\end{equation*}


\subsection{Hypergeometric orthogonal polynomials}	
The $n$th-order hypergeometric orthogonal polynomials ($q$-Krawtchouk $K_{n}(q^{-x};p,N;q)$ with $0<q<1$, $p>0$, $q$-Hahn $Q_{n}(q^{-x};\alpha,\beta,N;q)$ with $0<\alpha q<1$, $0<\beta q<1$, $q$-Meixner $M_{n}(q^{-x};b,c;q)$ with $0<bq<1$, $c>0$ and $q$-Charlier $C_{n}(q^{-x};a;q)$ with with $a>0$~\cite{Koekoek2}), 
are those polynomials that satisfy the orthogonality condition
\begin{equation*}
\sum_{x=0}^NP_{m}(q^{-x})P_{n}(q^{-x})w^{\lambda}(x)=d^2_{n}(\lambda)\delta_{m,n},\quad \lambda=6,\ldots,9,
\end{equation*}
where $N\in\Z^{+}$, $0\leq m,n\leq N$. Here,  the weight function $w^{\lambda}(\cdot)$ and the squared norm $d^2_{n}(\lambda)$, $\lambda=6,\ldots,9$, corresponding to these polynomials, respectively, are given in the Tables \ref{tableqKqH2}--\ref{tableqCqM2}.

On the other hand, these polynomials satisfy the following recurrence relation
\begin{equation*}
(q^{-x}-1)P_{n}(q^{-x}) =E_nP_{n+1}(q^{-x})-[E_n+F_n]P_{n}(q^{-x})+F_nP_{n-1}(q^{-x}),\quad n\geq 1,
\end{equation*}
where the coefficients $E_n$ and $F_n$ are given in the Tables \ref{tableqKqH2}--\ref{tableqCqM2}. This result is used to calculate higher-order hypergeometric orthogonal polynomials.

To avoid numerical fluctuations, we rescale these polynomials in order to obtain a more stable version. A normalized version can be defined by $\K_{n}^{\lambda,N}(x)$, with $\lambda=6,\ldots,9$, see Tables \ref{tableqKqH2}--\ref{tableqCqM2}.

In continuation we define $\K_{n}^{10,N}(x)$ corresponding to the discrete cosine transform
\begin{equation*}
\K_{n}^{10,N}(x)=\sigma(n)\cos\left(\fracd{\pi n(2x+1)}{2N} \right),\quad\sigma(n) =
\begin{cases}
\displaystyle \sqrt{1 / N} & \text{if} \qquad n = 0, \\
\displaystyle\sqrt{2 / N} & \text{otherwise.}%
\end{cases}
\end{equation*}

\begin{table}[tbph]
	\caption{Characterization of the $q$-Krawtchouk and $q$-Hahn polynomials.}
	\centering
	\renewcommand{\arraystretch}{2.0}
	\begin{tabular}{ccc} 
		\hline
		$P_n(q^{-x})$ & $K_{n}(q^{-x};p,N;q)$\: $(\lambda=6)$ 
		& $Q_{n}(q^{-x};\alpha,\beta,N;q)$\: $(\lambda=7)$ \\ \hline 
		$w^{\lambda}(x)$ & $(-p)^{-x}\frac{(q^{-N};q)_{x}}{(q;q)_x}$ & $\frac{(\alpha q,q^{-N};q)_x}{(q,\beta^{-1}q^{-N};q)_x}(\alpha\beta q)^{-x}$ \\ 
		$d^2_{n}(\lambda)$ & $\frac{(1+p)(q,-pq^{N+1};q)_n(-pq;q)_N(-pq^{-N})^n}{p^{N}(1+pq^{2n})(-p,q^{-N};q)_n}q^{n^2-\binom{N+1}{2}}$ & $\frac{(\alpha\beta q^2;q)_N(q,\alpha\beta q^{N+2},\beta q;q)_n(1-\alpha\beta q)(-\alpha q)^n}{(\beta q;q)_N(\alpha q,\alpha\beta q,q^{-N};q)_n(\alpha q)^N(1-\alpha\beta q^{2n+1})}q^{\binom{n}{2}-Nn}$ \\ 
		$E_n$ & $\frac{(1-q^{n-N})(1+pq^n)}{(1+pq^{2n})(1+pq^{2n+1})}$ & $\frac{(1-q^{n-N})(1-\alpha q^{n+1})(1-\alpha\beta q^{n+1})}{(1-\alpha\beta q^{2n+1})(1-\alpha\beta q^{2n+2})}$\\ 
		$F_n$ & $-pq^{2n-N-1}\frac{(1+pq^{n+N})(1-q^n)}{(1+pq^{2n-1})(1+pq^{2n})}$ & $-\frac{\alpha q^{n-N}(1-q^n)(1-\alpha\beta q^{n+N+1})(1-\beta q^n)}{(1-\alpha\beta q^{2n})(1-\alpha\beta q^{2n+1})}$\\ 
		$\K_{n}^{\lambda,N}(x)$ & $K_{n}(q^{-x};p,N;q)\sqrt{\frac{w^{6}(x)}{d^2_{n}(6)}}$ & $Q_{n}(q^{-x};\alpha,\beta,N;q)\sqrt{\frac{w^{7}(x)}{d^2_{n}(7)}}$ \\ \hline
	\end{tabular}
	\label{tableqKqH2}
\end{table}

\begin{table}[tbph]
	\caption{Characterization of the $q$-Charlier and $q$-Meixner polynomials.}
	\centering
	\renewcommand{\arraystretch}{2.0}
	\begin{tabular}{ccc} 
		\hline
		$P_n(q^{-x})$ & \quad $C_{n}(q^{-x};a;q)$\: $(\lambda=8)$ & $M_{n}(q^{-x};b,c;q)$\: $(\lambda=9)$\\ \hline
		$w^{\lambda}(x)$ & $\frac{a^x}{(q;q)_x}q^{\binom{x}{2}}$ & $\frac{(bq;q)_x}{(q,-bcq;q)_x}c^xq^{\binom{x}{2}}$\\ 
		$d^2_{n}(\lambda)$ & $q^{-n}(-a;q)_{\infty}(-a^{-1}q,q;q)_n$ & $\frac{(-c;q)_{\infty}}{(-bcq;q)_{\infty}}\frac{(q,-c^{-1}q;q)_n}{(bq;q)_n}q^{-n}$\\
		$E_n$ & $aq^{-2n-1}$ & $\frac{c(1-bq^{n+1})}{q^{2n+1}}$\\
		$F_n$ & $\frac{(1-q^{n})(a+q^{n})}{q^{2n}}$ & $\frac{(1-q^{n})(c+q^{n})}{q^{2n}}$ \\
		$\K_{n}^{\lambda,N}(x)$ & $C_{n}(q^{-x};a;q)\sqrt{\frac{w^{1}(x)}{d^2_{n}(1)}}$ & $M_{n}(q^{-x};b,c;q)\sqrt{\frac{w^{2}(x)}{d^2_{n}(2)}}$\\ \hline
	\end{tabular}
	\label{tableqCqM2}
\end{table}

\subsection{Orthogonal moments} 
Let $\C$ denote the cover image and let $(B_{i,j})$ be a block of $N\times N$ bytes of $\C$, with $i,j=0,\ldots,N-1$. Let $(\mathcal{B}^{\lambda,\xi}_{m,n}) $ be the corresponding block of $N\times N$ of the orthogonal moments of $(m+n)$-th order (direct moment transform), with $1\leq\lambda,\xi\leq 10$ and $m,n=0,\ldots,N-1$. The relationship between $\mathcal{B}^{\lambda,\xi}_{m,n}$ and its inverse $B^{\lambda,\xi}_{i,j}\equiv B_{i,j}$ (inverse moment transform) is given by~\cite{YapPO}
\begin{equation}
\mathcal{B}^{\lambda,\xi}_{m,n}=\sum_{0\leq
	i,j\leq N-1} \Ker^{\lambda,\xi}_{m,n}(i,j)B_{i,j},\label{DMT}
\end{equation}
\begin{equation}
B^{\lambda,\xi}_{i,j}=\sum_{0\leq
	m,n\leq N-1}\mathcal{B}^{\lambda,\xi}_{m,n}\Ker^{\lambda,\xi}_{m,n}(i,j),\label{IMT} 
\end{equation}
where $\Ker^{\lambda,\xi}_{m,n}(x,y)=\K_{m}^{\lambda}(x)\K_{n}^{\xi}(y)$ with $1\leq\lambda,\xi\leq 10$. Notice that, $\mathcal{B}^{\lambda,\lambda}_{m,n}$ with $1\leq\lambda\leq 9$ represents the orthogonal moments of Krawtchouk (K), 
Tchebichef (T), Hahn (H), Charlier (C), Meixner (M), $q$-Krawtchouk (qK), $q$-Hahn (qH), $q$-Charlier (qC) and $q$-Meixner (qM) respectively, and $\mathcal{B}^{10,10}_{m,n}$ represents the DCT. In addition, $\mathcal{B}^{\lambda,\xi}_{m,n}$ with $\lambda\neq\xi$ represents the combinations of the previous cases (separable moments), some of them studied in~\cite{BatiouBZ,hmimid2018image,Sayyou,TsougenisPK,ZhuH2}. For example, $\mathcal{B}^{4,7}_{m,n}$ represents the Charlier-$q$-Hahn moments, which we denote it by (CqH).

The software implementation of \eqref{DMT}--\eqref{IMT} can be easier computed by matrix multiplications, 
\begin{align*}
(\mathcal{B}^{\lambda,\xi}_{m,n}) = \A (B^{\lambda,\xi}_{i,j}  )\B^T,
\qquad
(B^{\lambda,\xi}_{i,j}) = \A^T (\mathcal{B}^{\lambda,\xi}_{m,n})\B,
\end{align*}
respectively, where $\A = (\K_{j}^{\lambda}(i))_{0\leq i,j\leq N-1}$ and $\B = (\K_{j}^{\xi}(i))_{0\leq i,j\leq N-1}$.

\subsection{Generation of chaotic positions}

The Beta function
\begin{equation*}
\beta(x,p,q,\varphi_1,\varphi_2) = \begin{cases}
\left( \fracd{x-\varphi_1}{\varphi_c-\varphi_1}\right)^p\left( \fracd{\varphi_2-x}{\varphi_2-\varphi_c}\right)^q & \mbox{ if } x\in\left[\varphi_1,\varphi_2 \right], \\ \\ 0 & \mbox{otherwise,}
\end{cases}
\end{equation*}
where $p,q,\varphi_1,\varphi_2\in \R$ with $\varphi_1<\varphi_2$ and
\begin{equation*}
\varphi_c = \fracd{p\varphi_2+q\varphi_1}{p+q},
\end{equation*}
is used in neural networks 
because of its high flexibility and its universal approximation characteristics \cite{zahmoul2017image}. According to scientific literature, several authors \cite{valandar2018integer,walia2018robust,yadav2018chaotic} have proposed novel steganographic algorithms based on chaotic maps, which are nonlinear systems suitable to design secure embedding methods  \cite{saidi2017new}. Indeed, these systems are characterized by a pseudo random behavior and an high sensitivity to initial conditions and control, unpredictability, ergodicity, etc \cite{martinez2016steganographic}.

In this work we use the Beta chaotic map created by the authors of \cite{zahmoul2017image}, which is mathematically defined by 
\begin{equation*}
x_{n} = r\beta(x_{n-1},p,q,\varphi_1,\varphi_2),\quad n\geq 1,
\end{equation*}
where $p=b_1+c_1a$ and $q=b_2+c_2a$, being $b_1,c_1,b_2$ and $c_2$ adequately chosen constants. The  parameter $r$, which is multiplied with the chaotic map, has the role to control the amplitude of the Beta map, and $a$ denotes the bifurcation parameter \cite{zahmoul2017image}. Thus, the chaotic positions can be generated by the Algorithm \ref{Chao_Pos}, 
and uses the following notations.

\begin{itemize}
	\item[\checkmark] card$(A)$ the cardinality of the set $A$ (number of elements of the set).
	
	\item[\checkmark] The function {\tt Reduce} returns the array without repeated elements.
	
	\item[\checkmark] $||$ concatenation.
	
	\item[\checkmark] $L\setminus\tau$ the set difference of $L$ and $\tau$.
\end{itemize}
Algorithm \ref{Chao_Pos} which calls and itself recursively and Algorithm \ref{BetaCMap}.

\begin{algorithm}[ht]
	\caption{$\overline{\beta}(x_0,n,r,a,b_1,c_1,b_2,c_2,\varphi_1,\varphi_2)$}
	\label{BetaCMap}
	\begin{algorithmic}
		\STATE \text{{\bf Input:}} 	$x_0,n,r,a,b_1,c_1,b_2,c_2,\varphi_1,\varphi_2$.
		\STATE \text{{\bf Output:}} $\left\lbrace x_1,\ldots,x_n \right\rbrace $.
		
		\STATE $\bullet$ $p\gets b_1+c_1a$;
		\STATE $\bullet$ $q\gets b_2+c_2a$;
		
		\FOR{$i=1$; $i\leq n$}
		\STATE $x_0\gets r\beta(x_0,p,q,\varphi_1,\varphi_2)$;
		\STATE $\triangleleft$ $x_i\gets\mbox{floor}(\mbox{mod}(10^{14}x_0),n)$;
		\ENDFOR
	\end{algorithmic} 
\end{algorithm}

\begin{algorithm}[ht]
	\caption{$\mbox{Chaotic-Positions\,}(x_0,L,r,a,b_1,c_1,b_2,c_2,\varphi_1,\varphi_2)$}
	\label{Chao_Pos}
	\begin{algorithmic}		
		
		\STATE \text{{\bf Input:}} 	$x_0,L,r,a,b_1,c_1,b_2,c_2,\varphi_1,\varphi_2$.
		\STATE \text{{\bf Output:}} $\rho=\{\rho_1,\ldots,\rho_{\mbox{card}(L)} \} $.

		\STATE $\bullet$ $\tau\gets\text{Reduce}(\overline{\beta}(x_0,\mbox{card}(L),r,a,b_1,c_1,b_2,c_2,\varphi_1,\varphi_2))$;
		\IF{$\mbox{card}(\tau)==1$}
		\STATE $\rho\gets L$;
		\ELSIF{$\mbox{card}(\tau)==\mbox{card}(L)$}
		\STATE $\rho\gets \tau$;
		\ELSE
		\STATE $\rho\gets\tau \; || \;  \mbox{Chaotic-Positions\,}(x_0,L\setminus\tau,r,a,b_1,c_1,b_2,c_2,\varphi_1,\varphi_2) $;
		\ENDIF
	\end{algorithmic} 
\end{algorithm}

\subsection{Key expansion of 2560 bits} 

There are two types of cryptography techniques, namely private key and public key cryptography. Public key cryptography is an asymmetric cryptography technique which encrypts the message with a private key and decrypts it with a public key. Private key cryptography is a symmetric cryptography technique which encrypts and decrypts a message with the same key. Advanced Encryption Standard (AES)~\cite{AESA} is a standard for the encryption of electronic data, which was accepted as FIPS standard by the National Institute of Standards and Technology (NIST) in November 2001. AES is a symmetric key block cipher, which means that the same key is used both by sender and receiver. It can encrypt and decrypt data blocks of 128 bits, using key size of 128, 192, and 256 bits. Moreover, most of the operations in the AES algorithm take place on bytes of data, which are represented in the field GF$(2^8)$, called the Galois Field. AES can be implemented on various platforms especially on small devices. It is carefully tested for many security applications.

\begin{algorithm}[ht]
\caption{Key Expansion}
\label{alg:keyexpansion}
\begin{algorithmic}
	\STATE \text{{\bf Input:}} 	$\kappa=\kappa_0\cup\kappa_1\cup\cdots\cup\kappa_{15}$.
	\STATE \text{{\bf Output:}} $\mathcal{P}$.
	%
	%
	\FOR{$p=1$; $p<3$}
	\FOR{$i=0$; $i<4$}
	\STATE $\omega_i\gets(\kappa_{4i},\kappa_{4i+1},\kappa_{4i+2},\kappa_{4i+3})$;
	\ENDFOR
	\FOR{$i=4$; $i<44$} 
	\STATE $\tau\gets\omega_{i-1}$;
	\IF{$i\mod{4} == 0$}
	\STATE $\tau\gets \text{Subbytes}(\text{RotWord}(\tau))\oplus \text{Rcon}(i/4)$;
	\ENDIF
	\STATE $\omega_i\gets\omega_{i-4}\oplus\tau$;
	\ENDFOR
	\STATE $\omega\gets\omega_{4}\cup\omega_{5}\cdots\cup\omega_{43}$;
	\IF{$p == 1$}
	\STATE $\vartheta\gets\omega$;
	\ELSE
	\STATE $\vartheta\gets\vartheta\cup\omega$;
	\ENDIF
	\STATE $\kappa\gets\omega_{40}\cup\omega_{41}\cup\omega_{42}\cup\omega_{43}$;
	\ENDFOR
	\STATE  $\triangleleft$ $\mathcal{P}\gets\text{byte2bin}(\vartheta)$;
\end{algorithmic} 
\end{algorithm}
In this paper we use the same steps as those developed for the AES to generate the key expansion of 2560 bits. From an initial key $\kappa=\kappa_0\cup\kappa_1\cup\cdots\cup\kappa_{15}$ of 16 bytes (128 bits), a binary sequence of 2560 bits is created as described in Algorithm \ref{alg:keyexpansion}.
The following list defines the used abbreviations.	
\begin{itemize}
		\item RotWord$(\cdot)$ takes a four-byte word and performs a cyclic permutation, i.e, RotWord$((a,b,c,d)) = (b,c,d,a)$.
		\item Subbytes$(\cdot)$ takes a four-byte input word and applies an S-box to each of the four bytes to produce an output word.
		\item Rcon$(\cdot)$ is a constant defined as:
		\begin{itemize}
			\item[\checkmark] Rcon$(j)=(R(j),0,0,0)$,
			\item[\checkmark] Each $R(j)$ is the element of Galois field GF$(2^8)$ corresponding to the value $x^{(j-1)}$ module $x^8+x^4+x^3+x+1$.
		\end{itemize}
		\item $\oplus$ is the exclusive OR operation, defined by:\\$0\oplus 0=1\oplus 1=0$ and $0\oplus 1=1\oplus 0=1$.
		\item byte2bin$(\cdot)$ converts a byte sequence to a binary sequence.
\end{itemize}

\section{Proposed Algorithm}

In this Section we propose a new steganographic algorithm. It is assumed that $\mathcal{L}$ is the length of the binary sequence of the secrete message $\M=\left\{ m_{\ell} \in \{ 0, 1 \} \text{: }1\leq \ell\leq \mathcal{L} \right\}$, where $m_{\ell}$ is a bit containing $0$ or $1$. 

\subsection{Permutation rule}

We denote by $\Per(\nu,\varpi)$ and $\Per^{-1}(\nu,\varpi)$ the following functions
\begin{algorithm}[H]
		\caption{$\Per(\nu,\varpi)$}
		 \label{Permutation}
		\begin{algorithmic}	
			\STATE \text{{\bf Input:}} 	$\nu$, $\varpi$.
			\STATE \text{{\bf Output:}} $\upsilon$.
			\STATE $\bullet$ $j=0$;	
			\FOR{$i=1$; $i\leq\text{length}(\varpi)$}
			\IF{$\varpi(i) == 1$}
			\STATE $j=j+1$;
			\STATE $\upsilon(j)\gets\nu(i)$;
			\ENDIF
			\ENDFOR
			\IF{$j \neq \text{length}(\varpi)$}
			\FOR{$i=1$; $i\leq\text{length}(\varpi)$}
			\IF{$\varpi(i) == 0$}
			\STATE $j=j+1$;
			\STATE $\upsilon(j)\gets\nu(i)$;
			\ENDIF
			\ENDFOR
			\ENDIF\vspace*{.45cm}
		\end{algorithmic} 
\end{algorithm}
\begin{algorithm}[ht!]
		\caption{$\Per^{-1}(\nu,\varpi)$}
		 \label{No-Permutation}
		\begin{algorithmic}	
			\STATE \text{{\bf Input:}} 	$\nu$, $\varpi$.
			\STATE \text{{\bf Output:}} $\upsilon$.
			\STATE $\bullet$ $j=0$;
			\STATE $\bullet$ $\psi\gets{\nu_1,\ldots,\nu_{|\varpi|}}$;	
			\FOR{$i=1$; $i\leq\text{length}(\varpi)$}
			\IF{$\varpi(i) == 1$}
			\STATE $j=j+1$;
			\STATE $\upsilon(i)\gets\psi(j)$;
			\ENDIF
			\ENDFOR
			\STATE $\bullet$ $j=0$;
			\STATE $\bullet$ $\psi\gets{\nu_{|\varpi|+1},\ldots,\nu_{\text{length}(\varpi)}}$;
			\FOR{$i=1$; $i\leq\text{length}(\varpi)$}
			\IF{$\varpi(i) == 0$}
			\STATE $j=j+1$;
			\STATE $\upsilon(i)\gets\psi(j)$;
			\ENDIF
			\ENDFOR
		\end{algorithmic} 
\end{algorithm}

Here, we denote by $|X|$ the number of bits of an array of bits $X$ which are equal to $1$.

\subsection{Details on quantification procedure and zig-zag scan} 
In the quantification procedure, the blocks of $8\times8$ bytes are quantified by 
\begin{equation}
\varTheta^{\lambda,\xi}_{u,v}=\text{round}\left(\fracd{\mathcal{B}^{\lambda,\xi}_{u,v}}{Q_{u,v}^{\mu}}\right) ,\quad 0\leq u,v\leq 7.\label{quantification}
\end{equation}
where
\begin{equation}
Q^{\mu}=\chi(\mu)\begin{pmatrix}
16 & 11 & 10 & 16 & 24 & 40 & 51 & 61 \\ 
12 & 12 & 14 & 19 & 26 & 58 & 60 & 55 \\ 
14 & 13 & 16 & 24 & 40 & 57 & 69 & 56 \\ 
14 & 17 & 22 & 29 & 51 & 87 & 80 & 62 \\ 
18 & 22 & 37 & 56 & 68 & 109 & 103 & 77 \\ 
24 & 35 & 55 & 64 & 81 & 104 & 113 & 92 \\ 
49 & 64 & 78 & 87 & 103 & 121 & 120 & 101 \\ 
72 & 92 & 95 & 98 & 112 & 100 & 103 & 99  %
\end{pmatrix}, \label{QM}
\end{equation}
with
$\displaystyle\chi(\mu) = \frac{100-\mu}{50}$, with $50<\mu<100$, see \cite{SoriaSCN}.

\begin{figure}[ht]
	\centering\includegraphics[scale=.45]{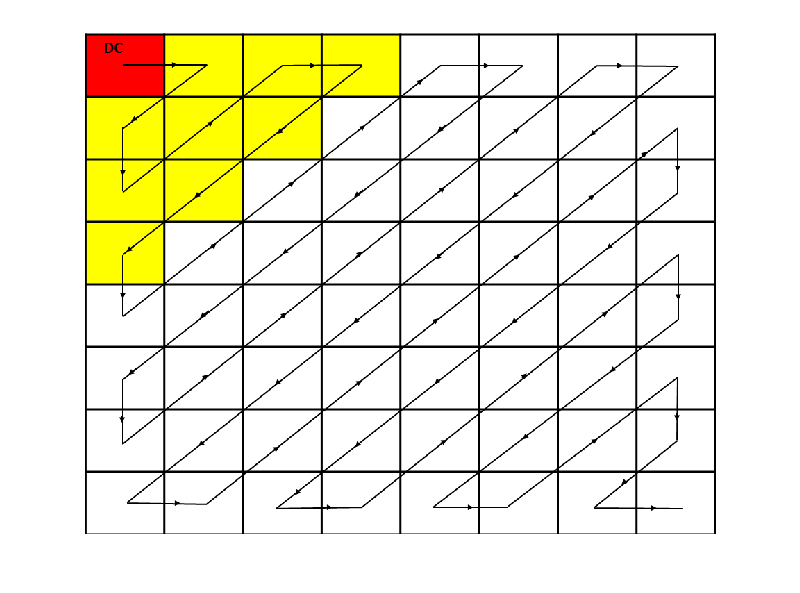}
	\caption[]{Zigzag order scan}
	\label{ZZagF}	
\end{figure}

The transformation of the matrix $\left( \varTheta^{\lambda,\xi}_{u,v}\right) $, with $0\leq u,v\leq 7$, to a vector $\nu^{\lambda,\xi} =\lbrace \nu_{i}^{\lambda,\xi} \text{: }1\leq i\leq 64\rbrace $ of length 64 is done by the zigzag order scan, see Figure \ref{ZZagF}, which aligns frequency coefficients in ascending order starting from frequency zero (DC coefficient) to high frequency components (AC coefficients), see~\cite{ZhuH2}. Indeed, the AC coefficients consist of three parts, 
those that occur at low, at middle and at high frequency, respectively,~\cite{SoriaSCN}. The non-zero AC coefficients occur at low and middle frequency, and perturbations to them do not affect the visual quality, whereas the zero AC coefficients occur usually at middle and high frequency, so modifications to them break the structure of continuous zeros and abrupt non-zero values give a hint of the existence of secret bits~\cite{Yu}.

\subsection{Embedding Algorithm}
The binary secret message $\M$ is inserted into the cover image by the embedding procedure described in Algorithm~\ref{alg:embedding}.

\noindent \textbf{Input}: Secret message $\M$, cover image $\C$, quality factor $\mu$, private key of 128 bits $\kappa$ and the initial conditions $x_0,a,b_1,c_1,b_2,c_2,\varphi_1,\varphi_2$, which
can also be adopted as a private key  jointly with the control parameter $r$.

\noindent \textbf{Output}: Stego image $\SM$.

\noindent \textbf{Procedure}: In the proposed algorithm, it is assumed that the emitter as well as the receiver hold the same system of private keys. Indeed, the emitter  generates the stego image from the private key of 128 bits and sends it trough an insecure channel to the receiver, which can extract the secret message from the aforementioned key. 

\begin{figure}[H]
	\centering\includegraphics[scale=.5]{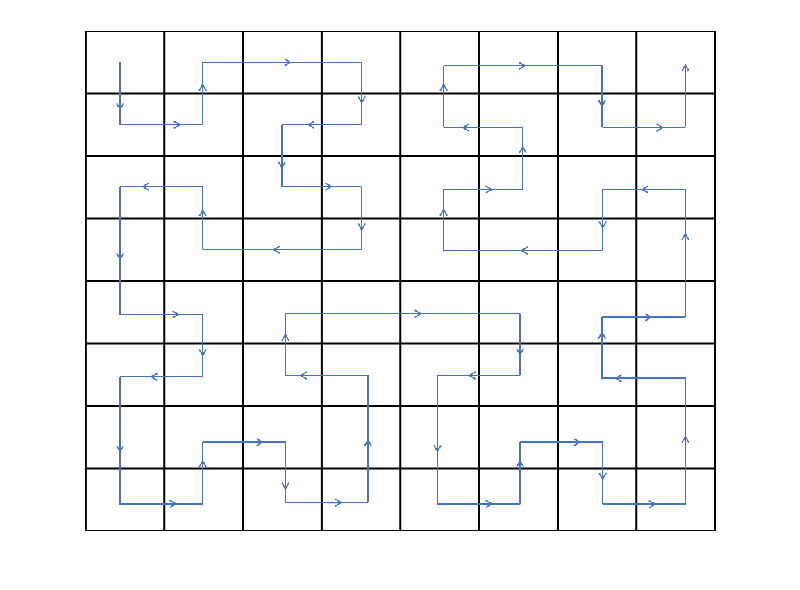}
	\caption[]{Hilbert order scan}
	\label{HSc}	
\end{figure}

The emitter generates the stego image $\SM$ according to Algorithm~\ref{alg:embedding}. Firstly, the proposed algorithm splits the cover image $\C = \bigcup_{k\in\mathcal{K}} B^{k}$ up into $\mbox{card}(\mathcal{K})$ non-overlapping blocks $B^{k}$ of $64\times 64$ bytes. Then, taking the chaotic positions (see Algorithm \ref{Chao_Pos}) into account, it divides each one of the blocks $B^{\rho_k}$ up into non-overlapping blocks of $8\times 8$ bytes, which are permuted by using the Hilbert order scan, see Figure \ref{HSc}. Next, it applies the direct moment transform \eqref{DMT} to each resultant block of $8\times 8$ bytes. Then, each one of these blocks is quantified according to the quantification matrix \eqref{QM}. Next, the zigzag scan is applied to the matrix of the quantized coefficients, see Figure \ref{ZZagF}, with the purpose to align frequency coefficients in ascending order. Afterwards, the first eight coefficients of high frequency are permuted by using the binary sequence $\mathcal{P}$ and the permutation rule $\Per(\cdot,\cdot)$. Thus, the secrete bits are embedded in the permuted coefficients. 
After the insertion of the secret message the back transformation is realized in reverse order: the permuted coefficients with the embedded secrete bits  are reorganized by using the binary sequence $\mathcal{P}$ and the permutation rule $\Per^{-1}(\cdot,\cdot)$, then by the zigzag scan the matrix of order $8$ is reconstructed, which afterwards is unquantified multiplying by the quantification matrix \eqref{QM}. Finally, the inverse moment transform \eqref{IMT} is applied in order to reconstruct the image, obtaining the expected stego image $\SM$.

For abbreviation we denote by
\begin{itemize}
	
	\item[\checkmark]   $\Delta(\eta)$ the function that reorganizes the vector $\eta$ of length 64 to a matrix of order $8$, taking into account the zigzag scan order \ref{ZZagF}.
	
	\item[\checkmark] R$(x,\beta)$ to the function that replaces the Least Significant Bit (LSB) of $x\in\N$ by $\beta\in\{0,1\}$, see \cite{SoriaSCN}.
\end{itemize}

\begin{algorithm}[H]
	\caption{Embedding Algorithm}
	\label{alg:embedding}
	\begin{algorithmic}	
		\STATE \text{{\bf Input:}} $\M,\C,\mu,x_0,a,b_1,c_1,b_2,c_2,\varphi_1,\varphi_2,r$;	
		\STATE \text{{\bf Output:}} $\SM$;
		\STATE $\triangleright$ Divide the cover image $\C = \bigcup_{k\in\mathcal{K}} B^{k}$ into $\mbox{card}(\mathcal{K})$ non-overlapping blocks $B^{k}$ of $64\times 64$ bytes; 
		
		\STATE $\triangleright$ $\{\rho_1,\ldots,\rho_{\mbox{card}(\mathcal{K})}\}\gets\mbox{Chaotic-Positions\,}(x_0,\left\lbrace 1,\ldots,\mbox{card}(\mathcal{K}) \right\rbrace ,r,a,b_1,c_1,b_2,c_2,\varphi_1,\varphi_2)$;
		
		\STATE $\triangleright$ $\{h_1,\ldots,h_{64}\}\gets$ Hilbert order scan, see Figure \ref{HSc};
		
		\STATE $\triangleright$ $\varsigma=\ell=0$;
		
		\FOR{$k\in\mathcal{K}$}
		
		\STATE $\triangleright$ Divide $B^{\rho_k} = \bigcup_{j\in\{1,\ldots,64\}} B^{\rho_{k},j}$ into $64$ non-overlapping blocks $B^{\rho_{k},j}$ of $8\times8$ bytes;
		
		\FOR{$j\in\{1,\ldots,64\}$}
		
		\STATE $\triangleright$ $\mathcal{B}^{\lambda,\xi,j}\gets B^{\rho_{k},h_j}$ : Calculate the direct moment transform coefficients $(\mathcal{B}^{\lambda,\xi,j})$ for $(B^{\rho_{k},h_j})$ according to~\eqref{DMT};
		
		\STATE $\triangleright$ $\varTheta^{\lambda,\xi,j} \gets \mathcal{B}^{\lambda,\xi,j}$		
		: Quantify $\mathcal{B}^{\lambda,\xi,j}$ according to~\eqref{quantification};
		
		\STATE $\triangleright$ $\nu^{j} \gets \varTheta^{\lambda,\xi,j}$ : Apply the zigzag scan, see Figure \ref{ZZagF};
		
		\STATE $\triangleright$ $\varsigma\gets \mbox{mod}(\varsigma,2560)+1$;
		
		\STATE $\triangleright$ $\mathfrak{a}^j\gets \Per(\{\nu^{j}_2,\ldots,\nu^{j}_9\},\{\mathcal{P}_{8(\varsigma-1)+1},\ldots,\mathcal{P}_{8\varsigma}\})$;
		
		\FOR{$q\in\{1,\ldots,8\}$}

        \IF{$\ell<\mathcal{L}$}
		\STATE $\triangleright$ $\ell\gets\ell + 1$;
		\IF{$\mathfrak{a}^j_q<0$}
		\STATE $\triangleright$ $\overline{\mathfrak{a}}_q^{j}\gets -$R$(|\mathfrak{a}^j_q|, m_{\ell})$;
		\ELSE
		\STATE $\triangleright$ $\overline{\mathfrak{a}}_q^{j}\gets $R$(\mathfrak{a}^j_q, m_{\ell})$;
		\ENDIF
		\ENDIF
		\ENDFOR
		
		\STATE $\triangleright$ $\{\nu^{j}_2,\ldots,\nu^{j}_9\}\gets \Per^{-1}(\overline{\mathfrak{a}}_k^{j},\{\mathcal{P}_{8(\varsigma-1)+1},\ldots,\mathcal{P}_{8\varsigma}\})$;
		
		\STATE $\triangleright$ $\overline{\varTheta}^{\lambda,\xi,j} \gets\Delta(\overline{\nu}^j)$;
		
		\STATE $\triangleright$ $\overline{\mathcal{B}}^{\lambda,\xi,j} \gets \overline{\varTheta}^{\lambda,\xi,j}$: Multiply the previous matrix by the quantification matrix $\left( \text{\ref{QM}}\right)$;
		
		\STATE  $\triangleright$ $\overline{B}^{\rho_{k},h_j} \gets \overline{\mathcal{B}}^{\lambda,\xi,j}$: Apply the inverse moment transform according to~\eqref{IMT};
		
		\ENDFOR

        \STATE $\triangleright$ $\overline{B}^{\rho_k}\gets \bigcup_j \overline{B}^{\rho_{k},j}$; 
          
		\ENDFOR
		\STATE $\triangleleft$ $\SM\gets\bigcup_k \overline{B}^{k}$;
	\end{algorithmic} 
\end{algorithm}

\subsection{Extracting Algorithm}
\noindent \textbf{Input}: Stego image $\SM$, quality factor $\mu$, private key of 128 bits $\kappa$ and the initial conditions $x_0,a,b_1,c_1,b_2,c_2,\varphi_1,\varphi_2$, which
can also be adopted as a private key  jointly with the parameter control $r$.

\noindent \textbf{Output}: Secret message $\mathcal{M}$.

\noindent \textbf{Procedure}: The receiver obtains the secret bits from Algorithm \ref{alg:xtracting}. \\

For abbreviation we denote the function that extracts LSB of $x\in\N$  by R$^{-1}(x)$, see \cite{SoriaSCN}.

\begin{algorithm}[H]
	\caption{Extracting Algorithm}
	\label{alg:xtracting}
	\begin{algorithmic}	
		\STATE \text{{\bf Input:}} $\SM,\mathcal{L},\mu,x_0,a,b_1,c_1,b_2,c_2,\varphi_1,\varphi_2,r$;	
		\STATE \text{{\bf Output:}} $\M$;
		\STATE $\triangleright$ Divide the stego image $\SM = \bigcup_{k\in\mathcal{K}} B^{k}$ into $\mbox{card}(\mathcal{K})$ non-overlapping blocks $B^{k}$ of $64\times 64$ bytes; 
		
		\STATE $\triangleright$ $\{\rho_1,\ldots,\rho_{\mbox{card}(\mathcal{K})}\}\gets\mbox{Chaotic-Positions\,}(x_0,\left\lbrace 1,\ldots,\mbox{card}(\mathcal{K}) \right\rbrace ,r,a,b_1,c_1,b_2,c_2,\varphi_1,\varphi_2)$;
		
		\STATE $\triangleright$ $\{h_1,\ldots,h_{64}\}\gets$ Hilbert order scan, see Figure \ref{HSc};
		
		\STATE $\triangleright$ $\varsigma=\ell=0$;
		
		\FOR{$k\in\mathcal{K}$}
		
		\STATE $\triangleright$ Divide $B^{\rho_k} = \bigcup_{j\in\{1,\ldots,64\}} B^{\rho_{k},j}$ into $64$ non-overlapping blocks $B^{\rho_{k},j}$ of $8\times8$ bytes;
		
		\FOR{$j\in\{1,\ldots,64\}$}
		
		\STATE $\triangleright$ $\mathcal{B}^{\lambda,\xi,j}\gets B^{\rho_{k},h_j}$ : Calculate the direct moment transform coefficients $(\mathcal{B}^{\lambda,\xi,j})$ for $(B^{\rho_{k},h_j})$ according to~\eqref{DMT};
		
		\STATE $\triangleright$ $\varTheta^{\lambda,\xi,j} \gets \mathcal{B}^{\lambda,\xi,j}$		
		: Quantify $\mathcal{B}^{\lambda,\xi,j}$ according to~\eqref{quantification};
		
		\STATE $\triangleright$ $\nu^{j} \gets \varTheta^{\lambda,\xi,j}$ : Apply the zigzag scan, see Figure \ref{ZZagF};
		
		\STATE $\triangleright$ $\varsigma\gets \mbox{mod}(\varsigma,2560)+1$;
		
		\STATE $\triangleright$ $\mathfrak{a}^j\gets \Per(\{\nu^{j}_2,\ldots,\nu^{j}_9\},\{\mathcal{P}_{8(\varsigma-1)+1},\ldots,\mathcal{P}_{8\varsigma}\})$;
		
		\FOR{$q\in\{1,\ldots,8\}$}
		
		\IF{$\ell<\mathcal{L}$}
		
		\STATE $\triangleright$ $\ell\gets\ell + 1$;

		\STATE $\triangleleft$ $m_{\ell}\gets $R$^{-1}(|\mathfrak{a}^j_q|)$;
		
		\ENDIF
		
		\ENDFOR
		
		\ENDFOR

		\ENDFOR
	\end{algorithmic} 
\end{algorithm}

\section{Results and discussion}

In this section the experimental results of the proposed algorithm are presented. The proposed algorithm is implemented in Python 3.7 for both Windows and Linux operating systems. 
The source codes will be made accessible in the complementary material.
For the experimental analysis several color images with size (512$\times$512) were collected from four different datasets: image dataset of 1500 RGB-BMP images, transformed from Caltech birds' dataset in JPEGC format \cite{AlJarrahM}, image dataset of 1500 RGB-BMP images, transformed from NRC dataset in TIFF format \cite{AlJarrahM}, image database \cite{Dataset_I} available in the University of Southern California and image dataset available at \cite{hlevkin}. For the experimental results several different randomly generated keys were used. We have tested our proposed algorithm by inserting a message of 98304 bits.

In addition, a comparison of the proposed method with respect to the methods proposed by Habib et al. (Hab. M.) \cite{habib2015enhancement}, Sahar (Sah. M.) \cite{el_rahman2016comparative}, Saidi et al. (Said. M.)~\cite{saidi2017new} and Chowdhuri et al. (Chow. M.) \cite{chowdhuri2018secured} is included in this section. The performance of the proposed approach has been studied using different kinds of statistical measures.

\subsection{Imperceptibility test}
The distortion level of the stego image with respect to its cover image in a steganographic system is measured in terms of Peak Signal to Noise Ratio (PSNR) \cite{MalathiPG}, which is calculated using the Mean Square Error (MSE). In addition, this measure is used to evaluate the invisibility of a secret message \cite{Atta2018} as well as the imperceptibility \cite{AAwad} and the visual quality \cite{Gaurav2018,NipanikarSI,ValandAB} of the stego image compared to the cover image, with decibel (db) as measurement unit. Higher PSNR indicates that the reconstruction of the image is of higher quality, \cite{DattaBK}. The PSNR is given by \cite{SoriaSCN}
\begin{equation*}
\text{PSNR}=10\log _{10}\left( \fracd{\Xi^{2}}{\text{MSE}}\right) ,
\end{equation*}
where
\begin{equation*}
\text{MSE}=(mn\rho)^{-1}\triplesum
\left\Vert
\mathcal{C}\left( \kk \right) -\mathcal{S}\left( \kk \right) \right\Vert ^{2},
\end{equation*}
and $\mathcal{C}$ and $\mathcal{S}$ are the cover image and the stego image respectively, of size $m\times n\times \rho$, with $\mathcal{C}, \mathcal{S} \in \{0, 1, \dots, \Xi \}$, and $\Xi=\text{max}(\text{max}(\mathcal{C}),\text{max}(\mathcal{S}))$.

The index set ${\kk} = (\ell_1, \ell_2, \ell_3)$ sums over the set
\begin{align*}
{\KK} = \{1, \dots, m \} \times \{ 1, \dots, n \} \times \{1, \dots, \rho \},
\end{align*}
where $\rho=1$ for gray scale images and  $\rho=3$ for color images of 24 bits.

\begin{figure}[H]
	\centering\includegraphics[scale=.438]{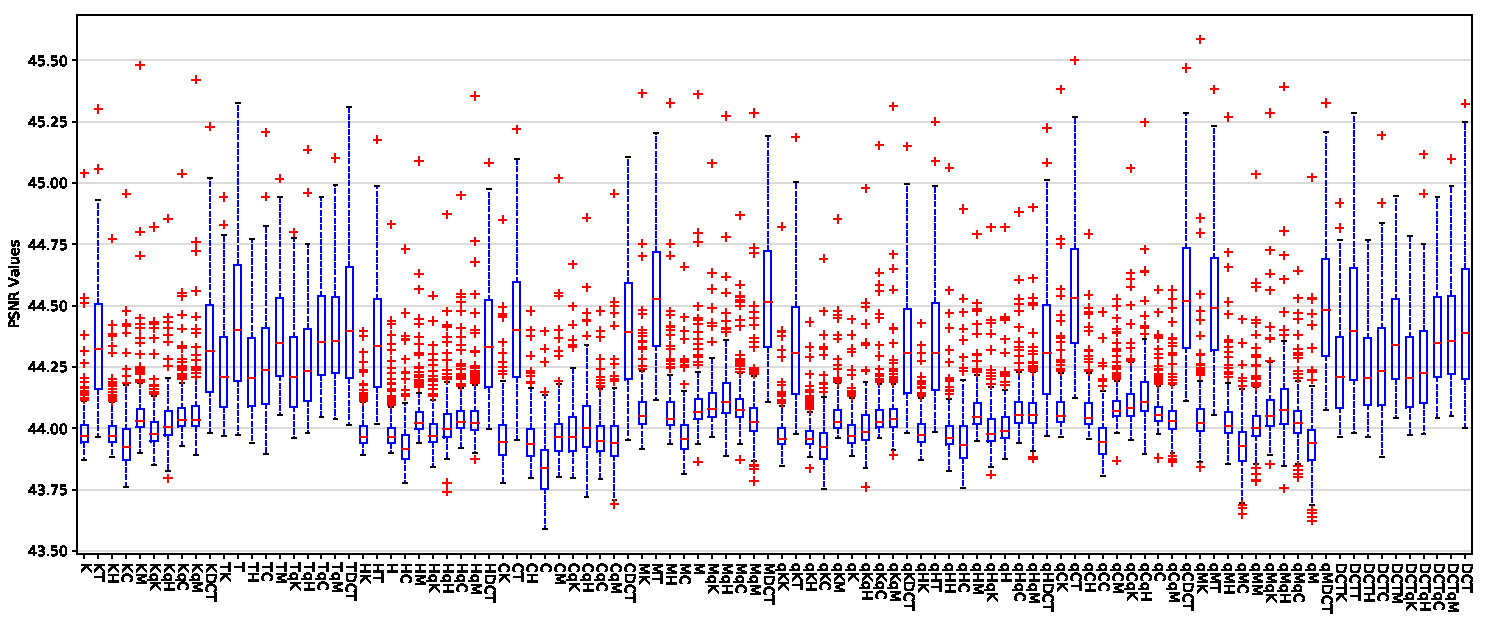}
	\caption[]{PSNR values corresponding to proposed method for the four datasets}
	\label{PSNR_datasets}	
\end{figure}

In the first experiment, we use the PSNR as a measure to evaluate the level of imperceptibility and distortion as well as to measure the different between cover and stego images. The experimental results showed that the proposed algorithm produced good quality stego images with good PSNR values, see Figure~\ref{PSNR_datasets}, which is in correspondence with the heuristic values of PSNR~\cite{RRoy,SoriaSCN}. Moreover, this experiment showed that for the four datasets, the results of imperceptibility corresponding to the proposed method for (T, TDCT, MT, MDCT, qCT, qCDCT, qMT, qMDCT and DCTT) were best to the obtained by proposed method for DCT transform, see Figure~\ref{PSNR_datasets}. 

\begin{figure}[H]
	\centering\includegraphics[scale=.48]{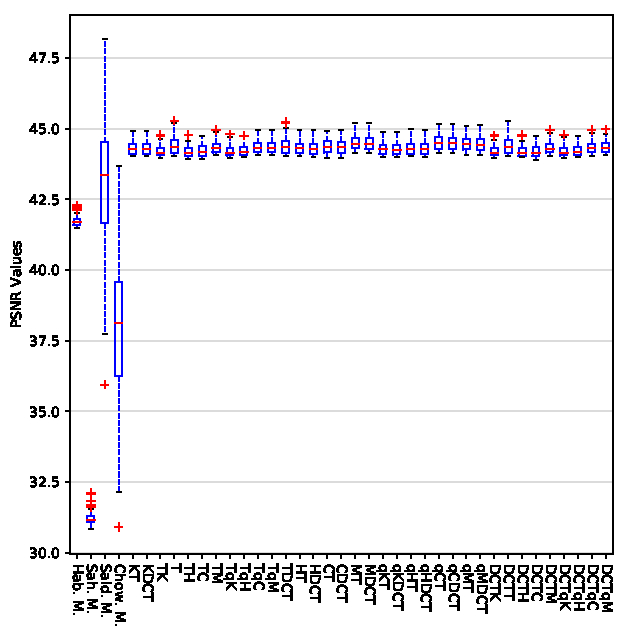}
	\centering\includegraphics[scale=.48]{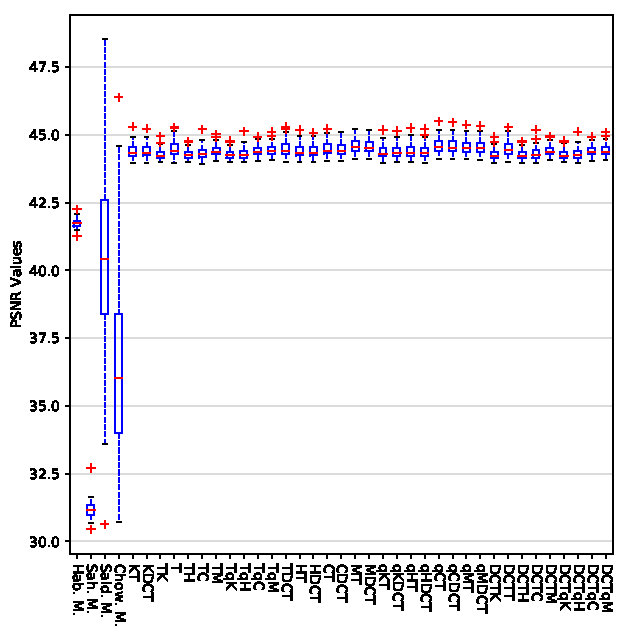}
	\centering\includegraphics[scale=.48]{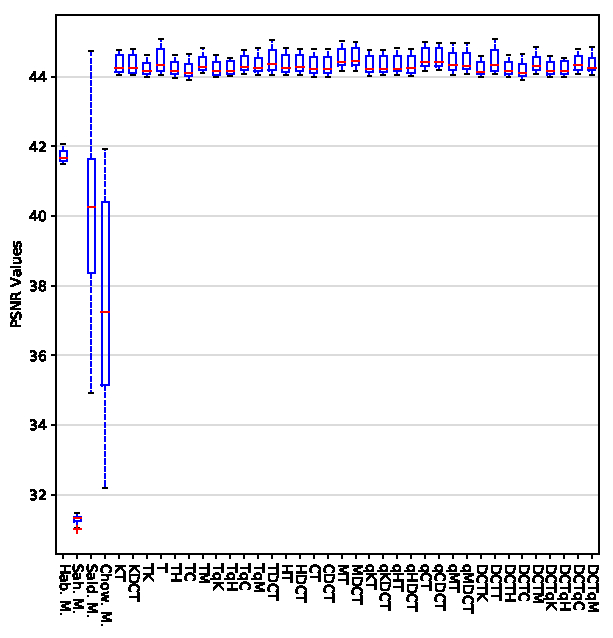}	
	\centering\includegraphics[scale=.48]{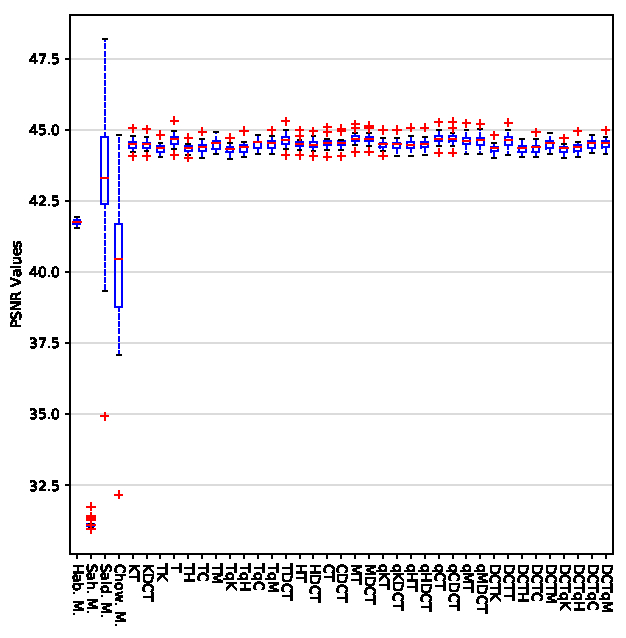}
	\caption[]{PSNR values. The first row contains the PSNR values corresponding to the first and second dataset while the second to third and fourth}
	\label{PSNR_DS-I-IV}	
\end{figure}

In the boxplots drawn in Figure~\ref{PSNR_DS-I-IV}, the horizontal axis represents the different methods that are compared, and the vertical axis represents the PSNR values. The upper and lower limit of the rectangle are the upper and lower quartiles ($Q_1$ and $Q_3$) of test results separately, and the difference between the upper and lower quartile is the quartile difference IQR. The red line in the rectangle is the median. The two black horizontal lines at $Q_3 + 1.5\text{IQR}$ and $Q_1-1.5\text{IQR}$ are the cut-off points for abnormal values, known as the internal limit. The data outside the internal limits is outliers and is represented by the red ‘+’. The Figure~\ref{PSNR_DS-I-IV} shows that the boxplots corresponding to proposed method are comparatively short, which suggests that overall PSNR values have a high level of agreement with each other. 

Moreover, for the four datasets the median of PSNR values corresponding to proposed method is greater in value than the median of PSNR values corresponding to Habib et al., Sahar, Saidi et al. and Chowdhuri et al. methods. In addition, of the four datasets is deduced that the 100$\%$ of PSNR values corresponding to proposed method is upper to PSNR values corresponding to Habib et al., Sahar and Chowdhuri et al. methods, while that the 75$\%$ of PSNR values corresponding to several orthogonal moments is upper to 75$\%$ of PSNR values corresponding to Saidi et al. method.

\subsection{Quality test}
\begin{figure}[H]
	\centering\includegraphics[scale=.48]{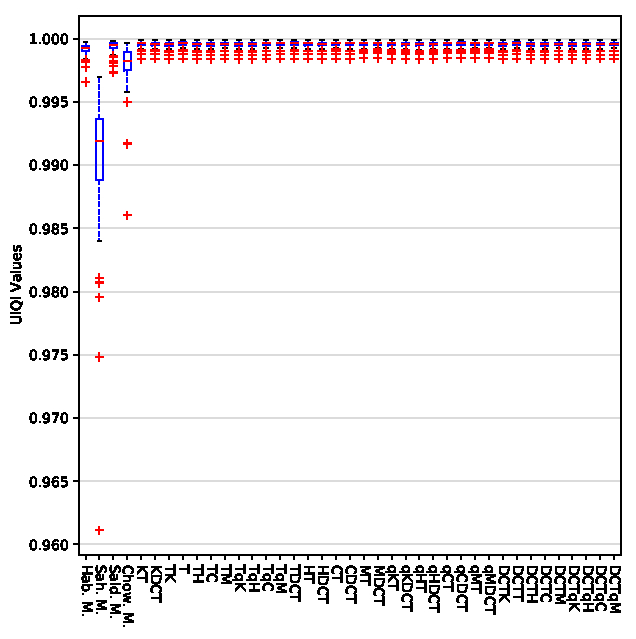}
	\centering\includegraphics[scale=.48]{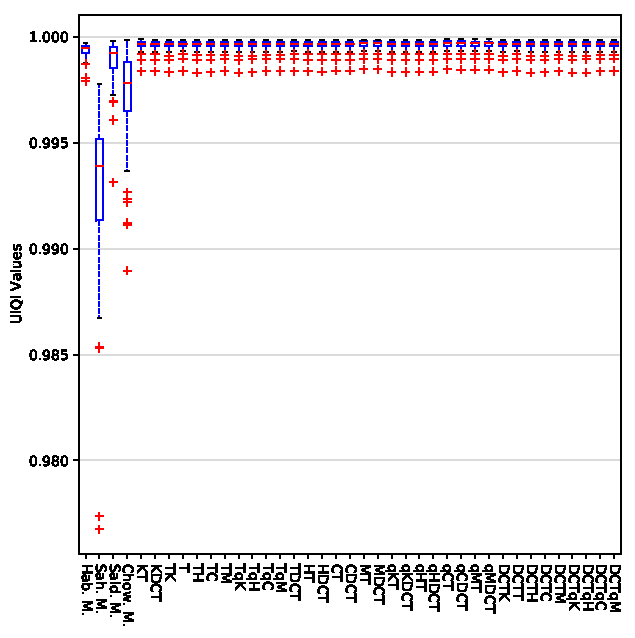}
	\centering\includegraphics[scale=.48]{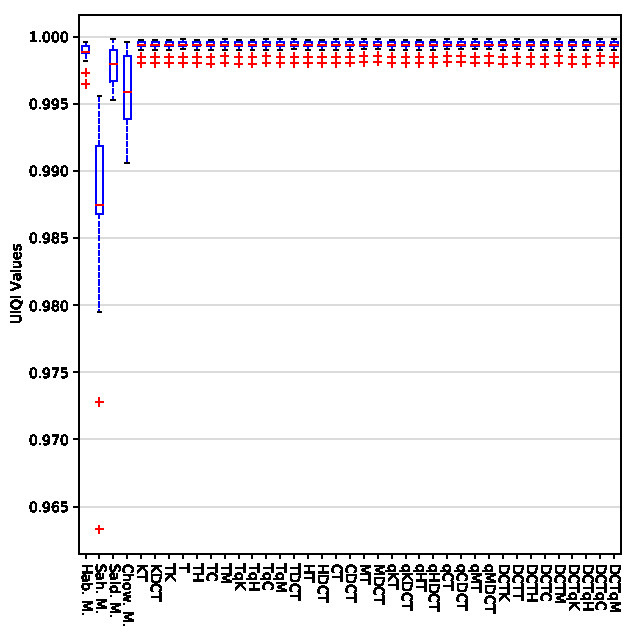}	
	\centering\includegraphics[scale=.48]{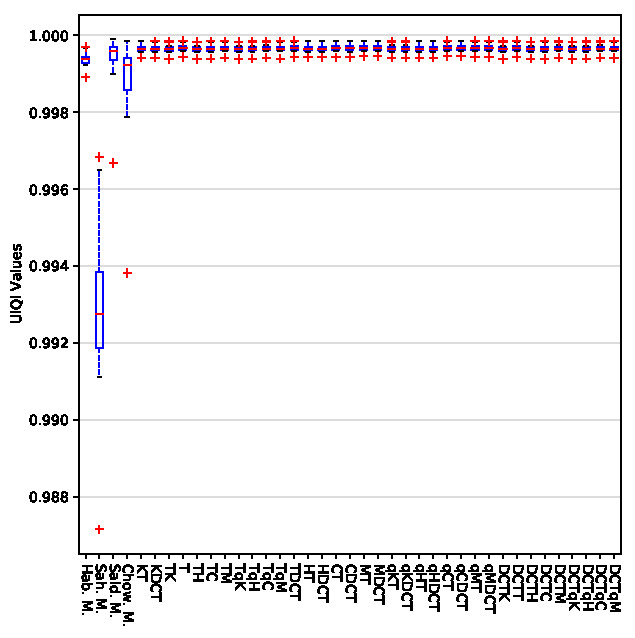}
	\caption[]{UIQI values. The first row contains the UIQI values corresponding to the first and second dataset while the second to third and fourth} 
	\label{UIQI-V}	
\end{figure}

Usually the image quality based on the Human Visual System (HVS) is measured by the Universal Image Quality Index (UIQI), which was proposed by Wang and Bovik in \cite{WangBo}. This measure is universal in the sense that it does not take the viewing conditions or the individual observer into account \cite{Bayra}. Moreover, it does not use traditional error summation methods \cite{ZhengYQ}. The dynamic range of UIQI is between -1 and 1. For identical images its value will be 1.
\begin{equation*}
\text{UIQI}=\fracd{4\sigma_{\mathcal{C}\mathcal{S}}}{\sigma_{\mathcal{C}}^2+\sigma_{\mathcal{S}}^2}\,\fracd{\overline{\mathcal{C}}\,\overline{\mathcal{S}}}{\overline{\mathcal{C}}^2+\overline{\mathcal{S}}^2},
\end{equation*}
where
\begin{equation*}
\overline{\mathcal{C}}=(mn\rho)^{-1}\triplesum
\mathcal{C}\left( \kk \right),
\end{equation*}
\begin{equation*}
\overline{\mathcal{S}}=(mn\rho)^{-1}\triplesum
\mathcal{S}\left( \kk \right),
\end{equation*}
\begin{equation*}
\sigma_{\mathcal{C}}^2=(mn\rho-1)^{-1}\triplesum
\left( \mathcal{C}\left( \kk \right)-\overline{\mathcal{C}}\right)^2,
\end{equation*}
\begin{equation*}
\sigma_{\mathcal{S}}^2=(mn\rho-1)^{-1}\triplesum
\left( \mathcal{S}\left( \kk \right)-\overline{\mathcal{S}}\right)^2,
\end{equation*}
\begin{equation*}
\sigma_{\mathcal{C}\mathcal{S}}=(mn\rho-1)^{-1}\triplesum
\left[\left( \mathcal{C}\left( \kk \right)-\overline{\mathcal{C}}\right)\left( \mathcal{S}\left( \kk \right)-\overline{\mathcal{S}}\right) \right] .
\end{equation*}
The second experiment shows that there are no significant differences between the cover and the stego images, since the UIQI values are close to unity. Additionally, in almost all the cases, the stego images obtained by the proposed method have major visual quality in comparison to the methods proposed by the other authors, see Figure~\ref{UIQI-V}.

\subsection{Similarity test}
Image fidelity is a measure that shows a consistent relationship with the quality perceived by the human visual perception. Moreover, measure it is a metric measure the similarity between the cover image $\mathcal{C}$ and the stego image $\mathcal{S}$ after insertion of the message \cite{SoriaSCN}. It is defined by~\cite{Khamrui,SenguptaM,SoriaSCN}
\begin{equation*}
\text{IF}=1-\fracsum{\displaystyle 
	\triplesum
	\left( \C \left(  \kk \right) - \SM \left( \kk  \right)\right) ^2}{\displaystyle
	\triplesum
	\C \left( \kk \right)^2} .
\end{equation*}
If the stego image is a close approximate of the cover image, then the value of IF would be close to unity.

\begin{figure}[H]
	\centering\includegraphics[scale=.48]{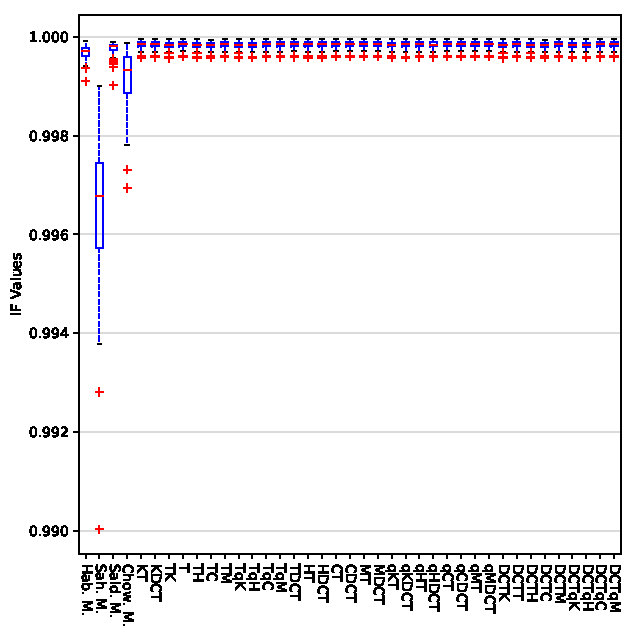}
	\centering\includegraphics[scale=.48]{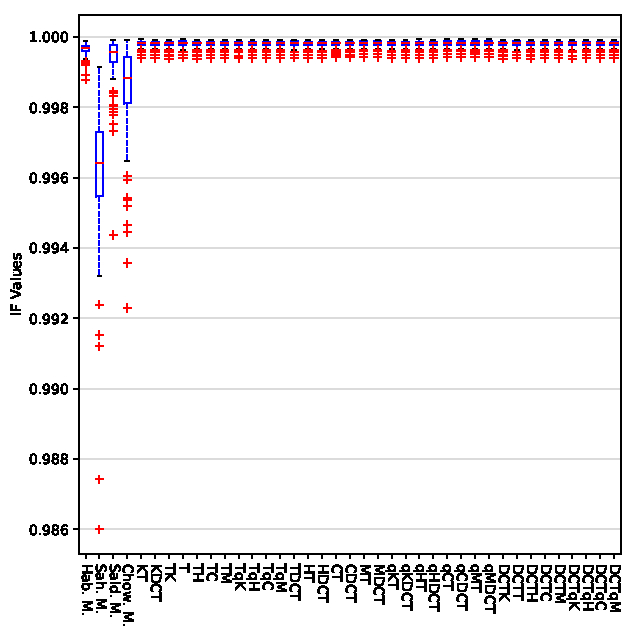}
	\centering\includegraphics[scale=.48]{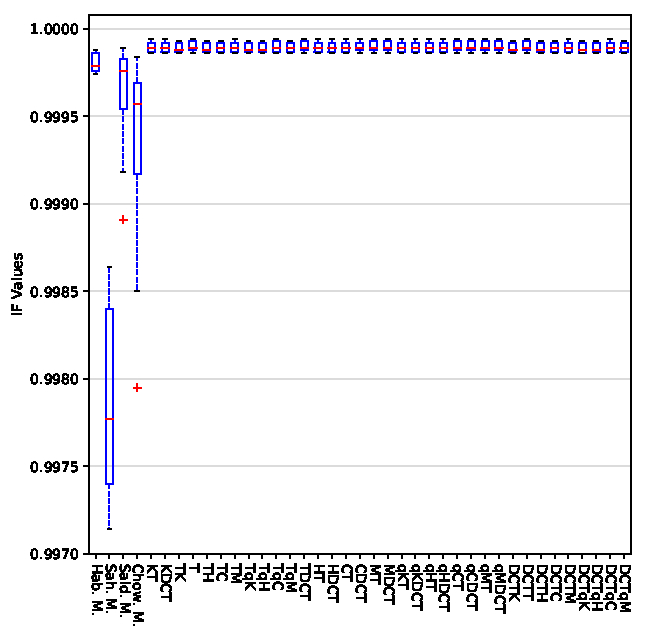}	
	\centering\includegraphics[scale=.48]{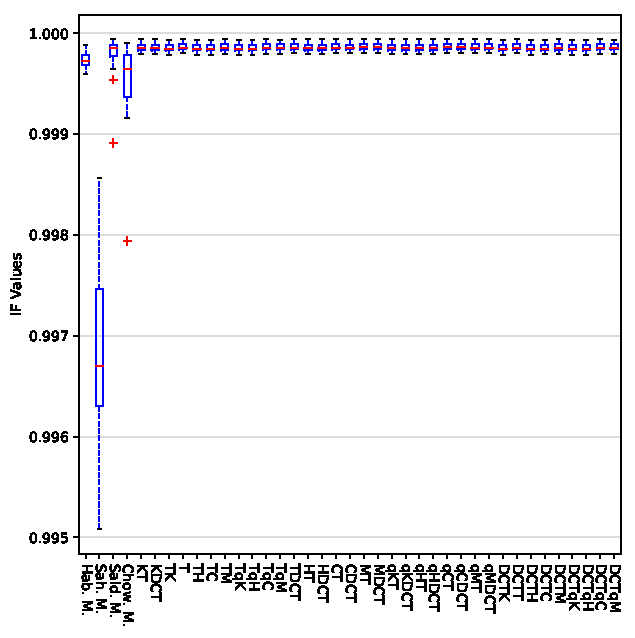}
	\caption[]{IF values. The first row contains the IF values corresponding to the first and second dataset while the second to third and fourth} 
	\label{IF-V}	
\end{figure}

In the third experiment, the values of IF were found for the four Datasets. It can be
observed that the IF values tend to one, see Figure \ref{IF-V}, which shows that is a high similarity between the cover image and the stego image after insertion of the secrete bits.

\subsection{Security test}	
The security of a steganographic system is defined in terms of the relative entropy
\begin{equation*}
\text{RE}\left( P_{\mathcal{C}}||P_{\mathcal{S}}\right) =\sum P_{\mathcal{C}}\left\vert \log \fracd{P_{\mathcal{C}}}{%
	P_{\mathcal{S}}}\right\vert,
\end{equation*}%
where $P_{\mathcal{C}}$ and $P_{\mathcal{S}}$ represent the distribution of cover and stego image, respectively. This statistical measure was proposed by Cachin in~\cite{Cachin,cachin2004information}. Moreover, a steganographic system is said to be
\begin{itemize}
	\item[\checkmark] $\varepsilon$-secure if $\text{RE}\left( P_{\mathcal{C}}||P_{\mathcal{S}}\right)\leq \varepsilon$,
	\item[\checkmark] perfectly secure if $\text{RE}\left( P_{\mathcal{C}}||P_{\mathcal{S}}\right) =0$.
\end{itemize}
Summing up, for the $\text{RE}\left( P_{\mathcal{C}}||P_{\mathcal{S}}\right)$, the closer the value is to 0, the higher the level of security.

\begin{figure}[H]
	\centering\includegraphics[scale=.43]{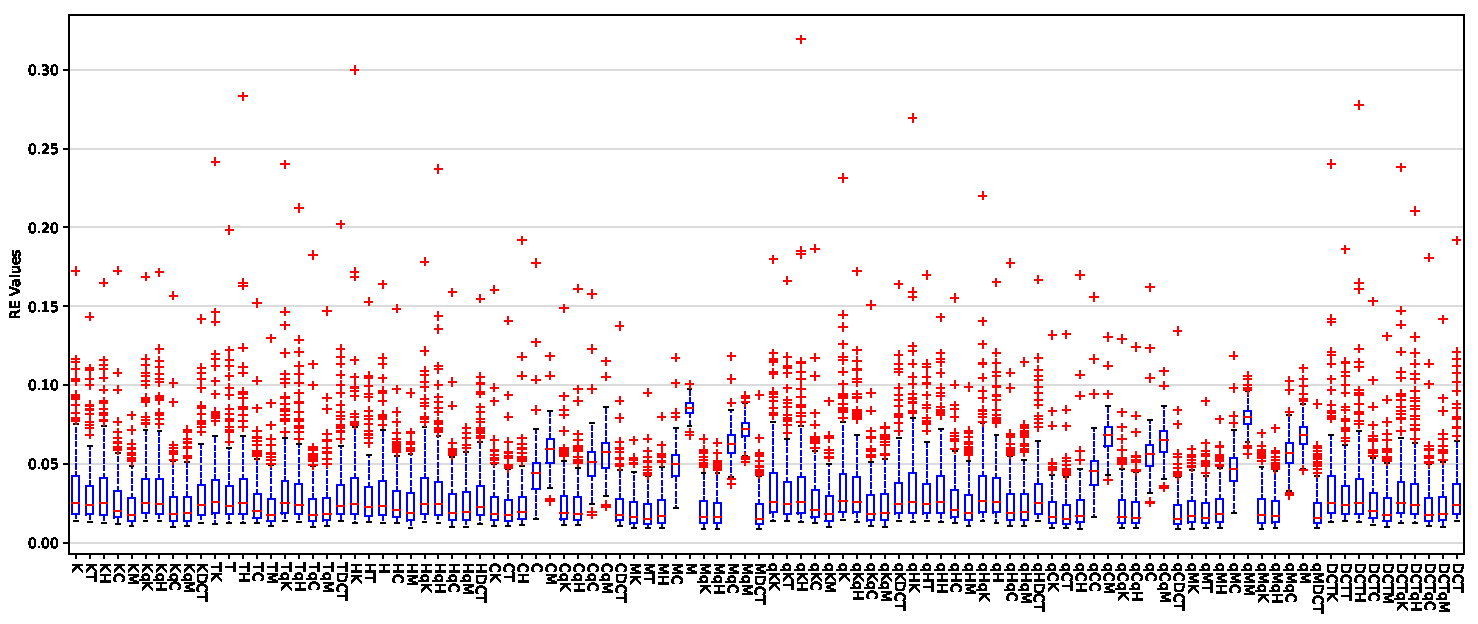}
	\caption[]{RE values corresponding to proposed method for the four datasets}
	\label{RE_DataSets_all}	
\end{figure}

\begin{figure}[H]
	\centering\includegraphics[scale=.48]{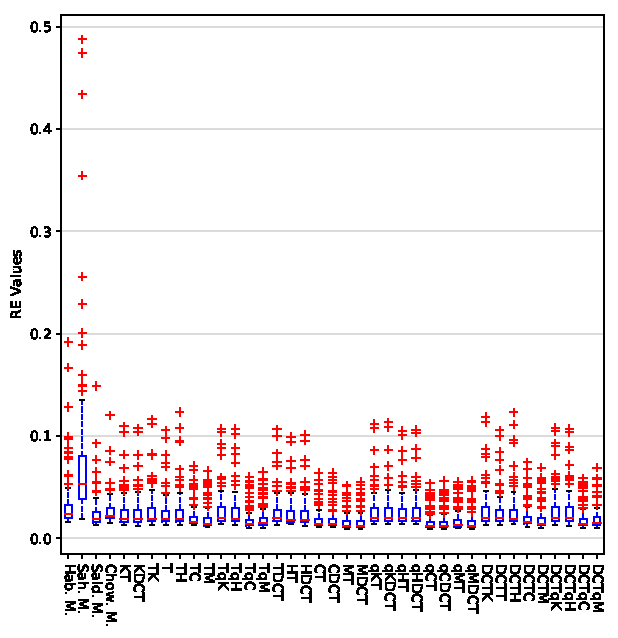}
	\centering\includegraphics[scale=.48]{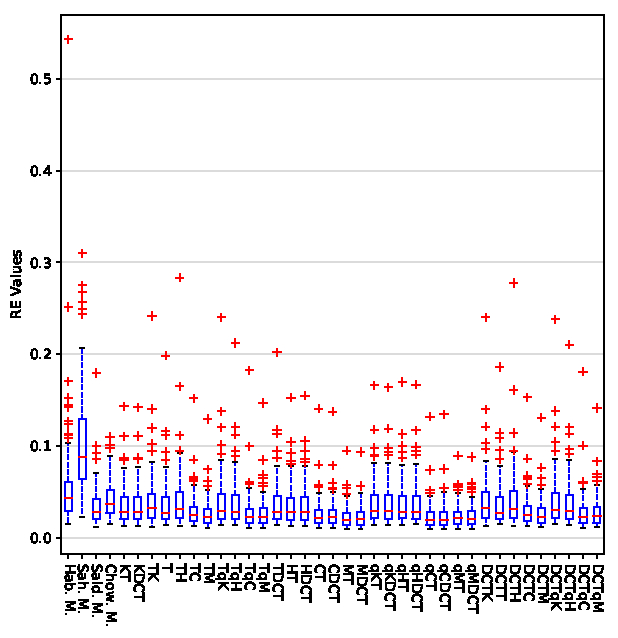}
	\centering\includegraphics[scale=.48]{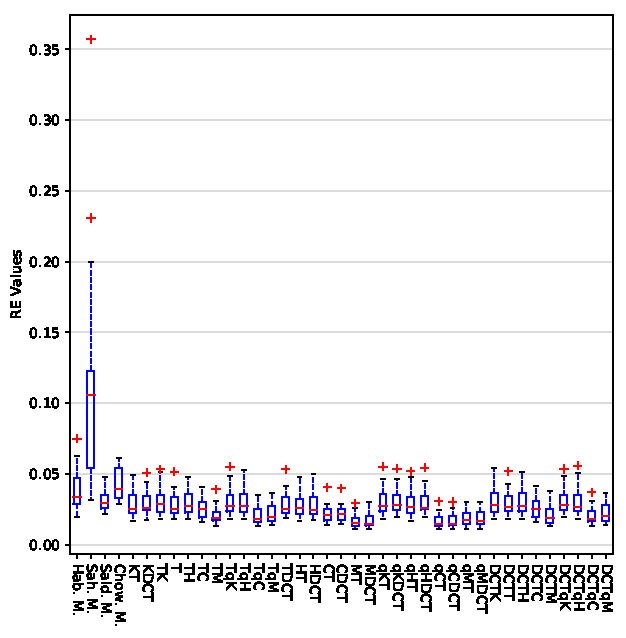}	
	\centering\includegraphics[scale=.48]{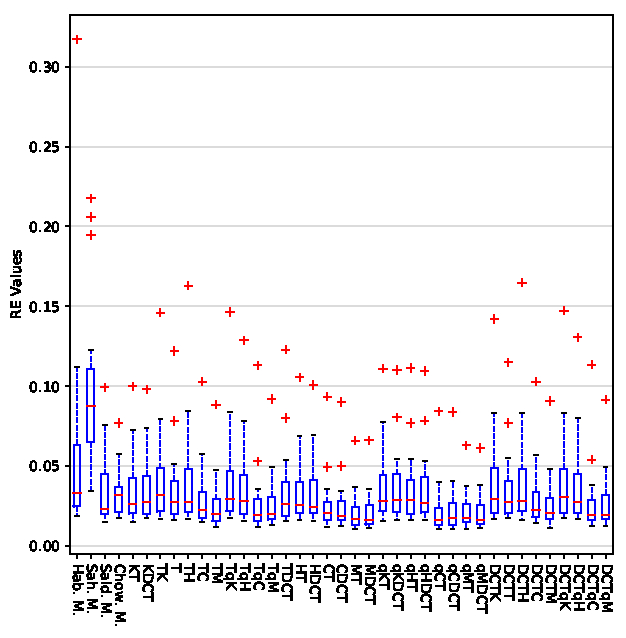}
	\caption[]{RE values. The first row contains the RE values corresponding to the first and second dataset while the second to third and fourth} 
	\label{RE_I-IV}	
\end{figure}

In the forth experiment we observe that the values of the relative entropy are close to zero, which affirms that the steganographic system obtained from the proposed algorithm is sufficiently secure, see Figures~\ref{RE_DataSets_all}--\ref{RE_I-IV}. Moreover, for the four datasets, the results of security corresponding to the proposed method for several orthogonal moments were best to the obtained by proposed method for DCT transform, see Figure~\ref{RE_DataSets_all}. And on the other hand, for the first dataset, the RE values obtained by the proposed method (TH, TM, TqC, TqM, CT, CDCT, MT, MDCT, qCT, qCDCT, qMT, qMDCT, DCTC, DCTM, DCTqC, DCTqM) are smaller in comparison to the obtained by Habib et al., Sahar, Saidi et al. and Chowdhuri et al., see Figure \ref{RE_I-IV}. Similar results are obtained for the other databases.

\subsubsection{Embedding capacity}
The performance of steganographic algorithms can be measured by two main criteria, embedding capacity and detectability. Thus novel steganographic algorithms are expected to increase the image capacity and the encryption strength of the message. The image capacity can be increased by adaptive strategies which decide where best to insert the message \cite{SoriaSCN}.

In steganography the {\em embedding capacity} is defined as the maximum number of bits that can be embedded in a given cover image. However, the {\em steganographic capacity} is the maximum number of bits that can be embedded in a given cover image with a negligible probability of detection by an adversary. Therefore, the embedding capacity is 
larger than the steganographic capacity~\cite{Sachd}.

\begin{figure}[ht]
	\centering\includegraphics[scale=.5]{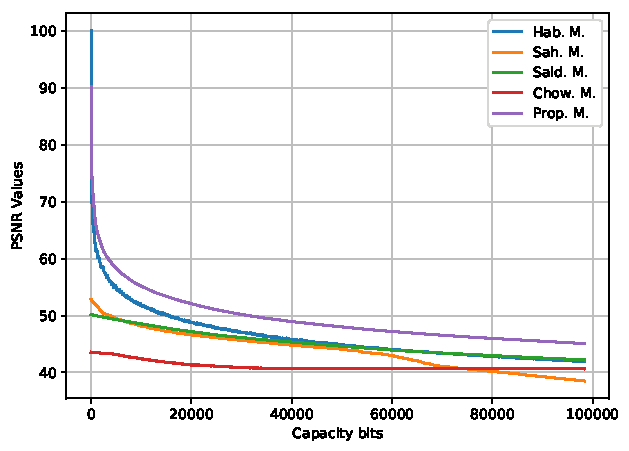}
	\centering\includegraphics[scale=.5]{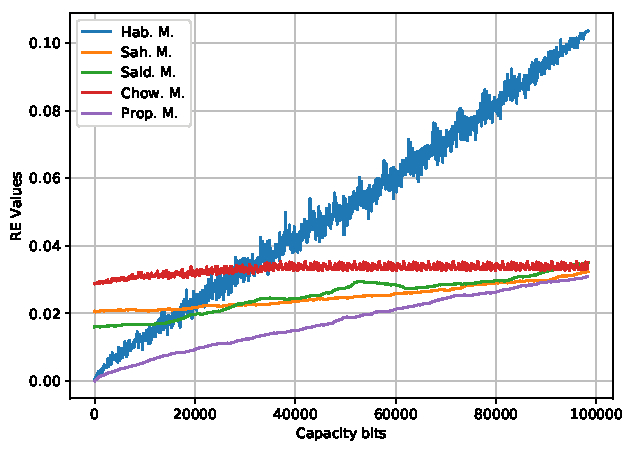}
	\caption[]{Embedding capacity} 
	\label{Cap}	
\end{figure}

In the fifth experiment we have tested embedding capacity through the PSNR values and the RE values. Here, we conclude that, the proposed approach can achieve high embedding capacity in comparison with the methods proposed by Habib et al., Sahar, Saidi et al. and Chowdhuri et al., keeping a high level of imperceptibility and security, see Figure \ref{Cap}.

\section*{Conclusions}

In this contribution, we propose a new steganographic algorithm which embeds a secrete message in the first eight AC coefficients. Here these coefficients are determined from some orthogonal polynomials and their combinations. Moreover, we use the Beta chaotic map to determine the order of the blocks where the secret bits will be inserted and we use a 128-bit private key to generate a key expansion of 2560 bits, with the propose to permute the first eight AC coefficients before the insertion. According to the analysis of PSNR, of UIQI values and of IF, it is demonstrated  that in the stego image there are no detectable anomalies to simple sight with respect to the cover image. Also, the obtained values for the relative entropy show that the steganographic system obtained by the proposed algorithm is sufficiently secure. In addition, the experimental results evidenced that the orthogonal moments MT, MDCT, qCT, qCDCT, qMT, qMDCT supply a higher level of imperceptibility keeping up an acceptable degree of security at the same time.


\subsection*{Acknowledgments}
The first wish to thank to ... 

The authors declare that there is no conflict of interest regarding the publication of this paper.











\end{document}